\documentclass[prl,showpacs,twocolumn,letterpaper,showkeys,
10pt,aps,superscriptaddress]{revtex4}
\usepackage{graphicx}
\usepackage{amsmath}
\usepackage{amssymb}

\usepackage[T1]{fontenc}
\usepackage{textcomp}
\usepackage{mathptmx}

\usepackage{marvosym}
\usepackage{dsfont}
\usepackage[latin1]{inputenc}

\begin{document}
\title{The emergence of atomic semifluxons in optical Josephson junctions} 

\author{M. Grupp} 
\author{W. P. Schleich} 
\affiliation{Universit{\"a}t Ulm, Institut f{\"u}r Quantenphysik and Center for Integrated Quantum Science and Technology (IQ$^{ST}$), D-89069 Ulm, Germany} 
\author{E. Goldobin}
\author{D. Koelle}
\author{R. Kleiner}
\affiliation{Physikalisches Institut II, Universit\"at T\"ubingen, D-72076 T\"ubingen, Germany} 
\author{R. Walser} 
\affiliation{Technische Universit\"at Darmstadt, Institut f\"ur Angewandte Physik, D-64289 Darmstadt, Germany} 
  \email{Reinhold.Walser@physik.tu-darmstadt.de}

\begin{abstract}
We propose to create pairs of semifluxons starting from a flat-phase state in long, optical $0$-$\pi$-$0$ Josephson junctions formed with internal electronic states of atomic  Bose-Einstein condensates.
In this optical system, we can dynamically tune the length $a$ of the $\pi$-junction, the detuning $\delta$ of the optical transition, or the strength $\Omega_0$ of the laser-coupling, to induce transitions from the flat-phase state to such a semifluxon-pair state. Similarly as in superconducting $0$-$\pi$-$0$ junctions, there are two, energetically degenerate semifluxon-pair states. A linear mean-field model with two internal electronic states explains this degeneracy and shows the distinct static field configuration in a phase-diagram of the junction parameters.  This optical system offers the possibility to dynamically create a coherent superposition of the distinct semifluxon-pair states and observe macroscopic quantum oscillation. 
 \end{abstract}
\date{2.11.2012}



\maketitle
The phenomenon of coupling coherent oscillators happens ubiquitously in mechanical and electrical systems \cite{Debnath2004}, in condensed matter physics \cite{Scott1969,barone71}, in nonlinear optics \cite{McCall1969,G_Lamb71} and in high energy physics \cite{Mandelstam1975}.
For many spatially extended fields, such as  laser pulses \cite{McCall1969,G_Lamb71}, superconducting Josephson junctions 
\cite{barone82,likharev86}
or atomic Bose-Einstein condensates \cite{smerzi1097,williams199,leggett401,oberthaler05,Kaurov2005,Kaurov2006,brand09,phillips11}, one can approximate the dynamics by the nonlinear sine-Gordon equation \cite{kivshar89}
for the real $2\pi$-periodic relative phase-field $\phi(t,x)$. 

A particularly interesting realization of such coupled oscillatory fields are superconducting $0$-$\pi$ Josephson junctions. Such junctions consist of segments where the ground state of the phase field $\phi(x)$ is either 0 or $\pi$. A $0$-$\pi$ junction can have semifluxons as local topological excitation which, in contrast to the well known fluxons, carry only half of the magnetic flux quantum $\Phi_0$ \cite{Bulaevskii:0-pi-LJJ,Xu:SF-shape,Kirtley:SF:HTSGB,Hilgenkamp:zigzag:SF}. 
The slightly more complicated $0$-$\pi$-$0$ junction can exhibit energy degenerate pairs of semifluxons as solutions. While the classical dynamics of semifluxons can be described well by sine-Gordon type equations \cite{Goldobin:SF-Shape,nappi04}, to address tunneling or 
macroscopic oscillations between the field configurations a requantization of the sine-Gordon phase field is required \cite{barone71,Mandelstam1975,goldobin05,vogel09,goldobin10}.

In the present Letter, we will demonstrate the appearance of the semifluxon-pair states starting from the flat-phase state in 
optical $0$-$\pi$-$0$ Josephson junctions implemented with two-level Bose-condensed atoms on a line $x$. Instead of the approximative sine-Gordon phase field, we will consider  a linear Schr{\"o}dinger equation  
as our mean-field model. As usual, we decompose the two complex field amplitudes $\psi_{\sigma}(t,x)=\sqrt{n_\sigma(t,x)}e^{i\phi_\sigma(t,x)}$
with densities $n_\sigma$ and phases $\phi_\sigma$.  
Then, we identify the phase difference $\phi\equiv \phi_e-\phi_g$ with the relative phase field of the sine-Gordon equation. 
For magnetically trapped $^{87}$Rb Bose-Einstein condensates, 
it is well known  that the relative phase is insensitive to density dependent energy shifts \cite{hall98,kuklov00,lewandowski02}. 
Thus, even a linear mean-field model exhibits the same energy-degenerate semifluxon states known from continuous \cite{goldobin05,vogel09,goldobin10}, or discrete \cite{walser07} quantum models. 

\begin{figure}[hT]
  \centering \includegraphics[angle=0,
  width=0.8\columnwidth]{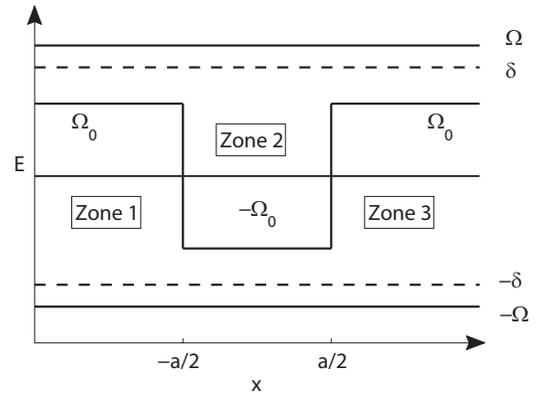}
  \caption{Spatial variation of frequencies in an optical $0$-$\pi$-$0$ Josephson junction for cold atoms: the Rabi-frequency $\Omega_0(x)$, the detuning $\delta$ and the generalized Rabi-frequencies $\pm \Omega$, which  are constant throughout the system. The junctions domains are localized at $\pm a/2$.}
  \label{fig1}
\end{figure}

We consider an atomic Bose-Einstein condensate with two internal states modeled 
by the Schr\"odinger equation $i\partial_t \psi(t,x)=H\psi(t,x)$,
where the atomic Hamiltonian operator
\begin{gather}
  H=-\partial_x^2+U, \quad
  U(x)=\begin{pmatrix}
    \delta&\Omega_0(x)\\\Omega^*_0(x)&-\delta
  \end{pmatrix},
\end{gather}
consists of kinetic energy and the dipole interaction $U(x)$ in presence of an external laser field \cite{footnote120911,schleich01}. 
Here, $\delta$ is the detuning of the laser
frequency from the atomic transition and the Rabi-frequency 
$\Omega_0(x)=\Omega_0 e^{i\theta(x)}$ is a
spatially dependent dipole coupling
derived from a laser field with constant modulus $|\Omega_0|$, but an abruptly jumping optical phase $\theta(x)$ as depicted in Fig.~{\ref{fig1}}. In general, this phase $\theta(x)$ could take on any value $\kappa$ \cite{goldobin10,vogel09}, but in order to model the $0$-$\pi$-$0$ Josephson junctions, we assume $\theta(x)=0,\pi$ or $0$, depending on zone 1,2 or 3. We emphasize that such a phase change can be realized by optical phase-imprinting techniques \cite{Dum1998a,dobrek99,denschlag00,strecker502}. 

The physics of the semi-fluxons in a single junction is given by an interplay between the mechanical motion as well as the coherent oscillation between the internal levels. Therefore, we introduce the dressed basis states $V=(V_+,V_-)$ of the local potential $U(x) V(\xi(x))=V(\xi(x))\Lambda$, with
\begin{align}
\label{dressed}
  V(\xi)\equiv \begin{pmatrix}
  \cos{\xi} & -\sin{\xi} \\ 
  \sin{\xi} &\phantom{-}\cos{\xi}
  \end{pmatrix}, \quad
  \tan{(\xi(x))}\equiv \frac{\Omega_0(x)}{\delta+\Omega}
\end{align}
and the generalized Rabi frequency $\Omega=\sqrt{|\Omega_0|^2+\delta^2}$ defines the
diagonal eigenvalue matrix $\Lambda=\mbox{diag}(\Omega,-\Omega)$. 

In a general Josephson junction array with $j$ zones, there are $j$ different $V_j$ eigen-matrices. However, it is a feature of this system that the eigenvalues $\Lambda=\Lambda_j$ are
all identical. Clearly, this fact is related to the light pressure
forces considered in atom trapping and cooling \cite{kazantsev90}. 
\begin{figure}[h]
  \centering \includegraphics[angle=-0,width=0.9\columnwidth]{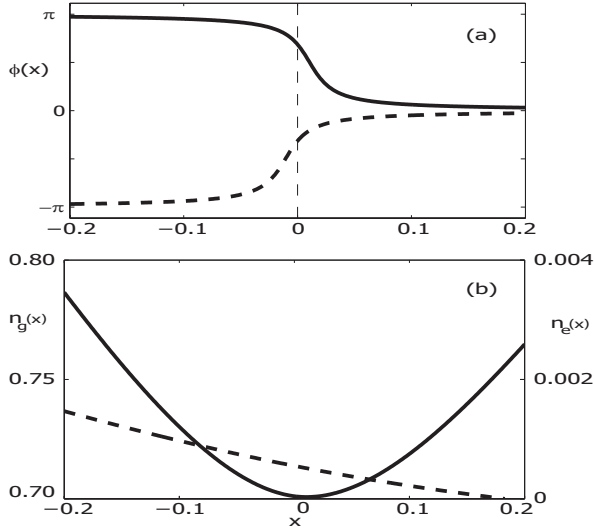}
 \caption{(a) Relative phase $\phi(x)\equiv \phi_e(x)-\phi_g(x)$, (b) ground state population $n_g(x)$ (left axis, dashed) and excited state populations $n_e(x)$ (right axis, solid) vs. position $x$ along the $0$-$\pi$ junction. In (a), we depict two energy-degenerate solutions, which have left (solid) and right (dashed) incoming plane wave asymptotics. The corresponding populations (b) are identical for both energy-degenerate solutions.  As parameters, we use $\Omega_0=1$, $\delta=3$ and a very low kinetic energy near the scattering threshold $E=-\Omega+k_-^2$ with $k_-^2=0.01$. The node of the excited state population $n_e$ at the location of the phase change resembles the physics of vortices in higher dimensions. \label{fig:0piwelle}}
\end{figure}

Before discussing the exact solution of the stationary Schrödinger equation $H\psi=E\psi $, it is important to find length scales.  By equating the energy of the internal motion ($\Omega$) with the mechanical energy ($1/a^2$), one can identify a characteristic length as $a_c=1/\sqrt{\Omega}$. In each
zone j, the physical solution 
\begin{align}
  {\psi(x)}=
  V (
  e^{i K x}{\psi^r} + e^{-i K x}{\psi}^l),
\end{align}
is a superposition of left- or right-propagating or attenuated waves
with upper $(+)$ and lower $(-)$ dressed state amplitudes $\psi^{h=r/l}=(\psi^h_+,\psi^h_-)$ according to Eq.~(\ref{dressed}). The compact matrix notation also extends to the wave number
$K\equiv\sqrt{E-\Lambda+i0^+}$, which is a diagonal matrix  with entries
$k_\pm\equiv \sqrt{E\mp\Omega+i0^+}$. For definiteness, we have shifted the square root into the upper complex plane and cut it along the negative real axis. This fact is
relevant as there are two distinct energy ranges: $-\Omega<E<\Omega$ and
$\Omega<E$. In the former case, $k_+$ has a positive imaginary part and $k_- >0$, while in the later case both $k_\pm >0$.

First, let us consider a single $0$-$\pi$ junction at $x_1=-a/2$. There, we have to match the solutions 
\begin{equation}
\begin{aligned}
\psi_1(x)&=V_1(e^{i K x}\psi_1^{in} +e^{-i K x}\psi_1^{out}),\\
\psi_2(x)&=V_2(e^{i K x}\psi_2^{out}+ e^{-i Kx}\psi_2^{in}
\end{aligned}	
\label{zeropi}
\end{equation}
in zone  1 and 2 requiring continuity $\psi_1(x_1)=\psi_2(x_1)$ and differentiability $\partial_x\psi_1(x_1)=\partial_x \psi_2(x_1)$. Quantitatively, we use the current density $j(x)\equiv \Im{\{\psi^{\dagger}\partial_x \psi\}}$ (imaginary part) to decide whether left-, or right-propagating waves in zone $j$ are counted as incoming  $\psi_j^{in}$ or outgoing  $\psi_j^{out}$ relative to the location of the junction at $x_j$. These definitions lead to a four-dimensional scattering matrix $S$ of the $0$-$\pi$ junction given by
\begin{align}
\label{smat}
    \varphi^{out}_{\beta}
    &=
    \sum_{\alpha}S_{\beta\alpha}
    \varphi^{in}_{\alpha}.
\end{align}
This relation quantifies the energy-dependent response of the system 
$\varphi^{out}=( \psi_{2}^{out},\psi_{1}^{out})$ to input signals $\varphi^{in}=( \psi_{1}^{in},\psi_{2}^{in})$ in the four different collision channels labeled by 
$\alpha\equiv(j=1,2;\sigma=\pm)$.  
In our Hamiltonian system, currents are conserved at all junctions. This implies the unitarity of the S-matrix, that is $S^{\dagger}gS=g,$ in all open channels with respect to the diagonal metric
$g=\Re\{\text{diag}(k_+,k_-,k_+,k_-)\}$.

This simple model can be solved analytically. In the energy range $-\Omega<E<\Omega$, the excited dressed state channels are energetically closed, i.\thinspace{}e., $\psi_i^{in/out}=(0,\psi_{i-}^{in/out})$
and the S-matrix between the open collision channels reads
\begin{gather}
\label{zeropis}
  \begin{pmatrix}
    \psi_{2-}^{out}\\ \psi_{1-}^{out}\end{pmatrix}=
  \begin{pmatrix}T&R\\
    R &T\end{pmatrix}
  \begin{pmatrix}\psi_{1-}^{in}\\ \psi_{2-}^{in}\end{pmatrix},
\end{gather}
where $R=(k_-^2-k_+^2)\sin^2(2\xi)/f$, $T=-4(k_-k_+)\cos(2\xi)/f$, 
and  $f=[k_-^2+6k_-k_++k_+^2-(k_--k_+)^2\cos(4\xi)]/2$.
The energy-dependent transmission $|T(E)|^2$ vanishes at $E=\pm \Omega$ and the simple maximum in between depends on the laser parameter $\xi(\Omega_0,\delta)$.  In Fig.~\ref{fig:0piwelle} a, we observe the expected $\pi$-phase flip between left- and right-asymptotic state amplitudes. Mathematically, this property is reflected in the negative sign in the transmission amplitude $T(E)$ defined in Eq.~(\ref{zeropis}). In the limit of very weak coupling this becomes $\lim_{\xi\rightarrow 0}T(E)=-1$. 

The classical part of the kinetic energy of the excited state population is proportional to $n_e(x)(\partial_x\phi_e)^2$. In order to minimize the energy change along the junction, a steep phase gradient has to be accompanied by a node in the excited state population as seen in  Fig.~\ref{fig:0piwelle} b. This is the same physical mechanism as the vanishing core density of two- or higher-dimensional superfluid vortices \cite{fetter2002}.   

In the atomic $0$-$\pi$-$0$ Josephson junction of Fig.~\ref{fig1}, we can now  find a qualitatively new feature: as before semifluxons occur on each location of the junctions, but only if the length $a$ of the $\pi$-junction  exceeds a characteristic value $a>a_{c}(\Omega)$, in analogy to superconductivity \cite{goldobin05}. Thus, a different motional topological state emerges.   
By generalizing the previous calculation, we add a wave-function for the middle zone and obtain the scattering solutions of the Schr{\"o}dinger equation by matching the pieces 
\begin{equation}
\begin{aligned}
  \psi_1(x)&=V_1(e^{i K x}\psi_1^{in} + e^{-i K x}\psi_1^{out}),\\
  \psi_2(x)&=V_2(e^{i K x}\psi_2^{r} + e^{-i K x}\psi_2^{l}),\\
  \psi_3(x)&=V_3(e^{i K x}\psi_3^{out}+ e^{-i K x}\psi_3^{in}),
\end{aligned}
\label{threezone}
\end{equation}
at  $x=\pm a/2$ as in the case of the $\pi$-junction. 
\begin{figure}[h]
  \centering 
  \includegraphics[angle=0,width=\columnwidth]{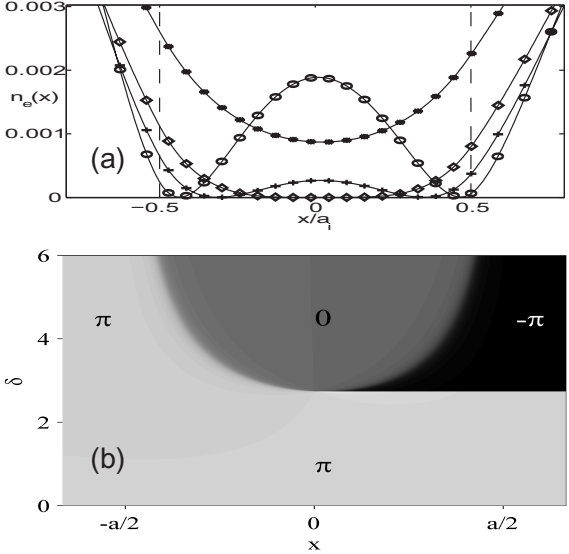}
  \caption{a) Excited state density $n_e(x;a_i)$ vs. a scaled length 
  coordinate $x/a_i$.   With the boundary condition of a left-incoming plane waves impinging on the 
  $0$-$\pi$-$0$ junction, we picked four different values $a_i$ for the length of the $\pi$-zone: 
  $a_1=0.4$ (*), $a_2=a_{c}\approx 0.571$ ($\lozenge$),
  $a_3=0.7$ (+) and $a_4=1$ (o). The laser parameters are 
  $\delta=3$, $\Omega_0=1$ and we have an energy near the 
  scattering threshold $E=-\Omega+k_-^2$ with $k_-^2=0.01$.
  b) Gray-scale density plot of the relative phase $\phi(x;\delta)$ versus 
  position and detuning $\delta$ with $a=0.6$. 
    By varying the detuning $\delta$, we can switch from the
    semifluxon regime ($\delta_{c}\approx 2.74$) to the flat phase state domain.}
  \label{fig:0pi0all}
\end{figure}

In Fig.~\ref{fig:0pi0all} a, we display the excited state density $n_e(x;a_i)$ as function of position in scaled coordinates $x/a_i$ for different values $a_i$ of the length of $\pi$-junction. We clearly see a qualitative change in the shape of the density when we increase the length to values above $a>a_c$. In the former case , the density is nonzero everywhere. By increasing the length of the junction to $a=a_c$ the density touches zero. A further increase of the length to $a>a_c$ leads to the formation of two nodes located $x=\pm a/2$ and a non-vanishing density in between. 

Already this static picture for the density suggests the formation of semifluxon pairs, when the length exceeds a critical length. 
However, this effect is also seen by directly studying the relative phase as a function of position and any system parameter $a$, $\Omega_0$ or $\delta$. If any one of the parameters varies while keeping the others fixed, we can observe the emergence of different quantum states in a two-dimensional phase-diagram. In particular, we have varied the laser detuning $\delta$ at fixed values for $a$ and $\Omega_0$ in Fig.~\ref{fig:0pi0all} b. This gray-scale density plot depicts the relative phase $\phi(x;\delta)$.
In essence, we find a semi-annular phase boundary limited by $|x|<a/2$ and $\delta>\delta_c$, which separates flat-phase domains ($\pi$) from semifluxon pair regions ($\pi$-$0$-$(-\pi)$).

So far, we have confined the discussion to low energy scattering $k_{-}^2=0.01$, as seen in Figs.~\ref{fig:0piwelle} and \ref{fig:0pi0all}. But this is no limitation for the experimental realizations of this system. Therefore, we also analyze the high energy scattering behavior with the S-matrix for the $0$-$\pi$-$0$  junction. It is defined as in Eq.~(\ref{smat}) and can be found explicitly from the solution of Eqs.~(\ref{threezone}).  
We only have to specify which out-going amplitudes $\varphi^{out}=( \psi_{3}^{out},\psi_{1}^{out})$ are connected by the S-matrix to 
the incoming amplitudes 
$\varphi^{in}=(\psi_{1}^{in},\psi_{3}^{in})$ in the four different 
collision channels of zone 1 and 3, now labeled by $\alpha=(j=1,3,\sigma=\pm)$.

\begin{figure}[h]
  \centering 
  \includegraphics[angle=0,width=1.5\columnwidth]{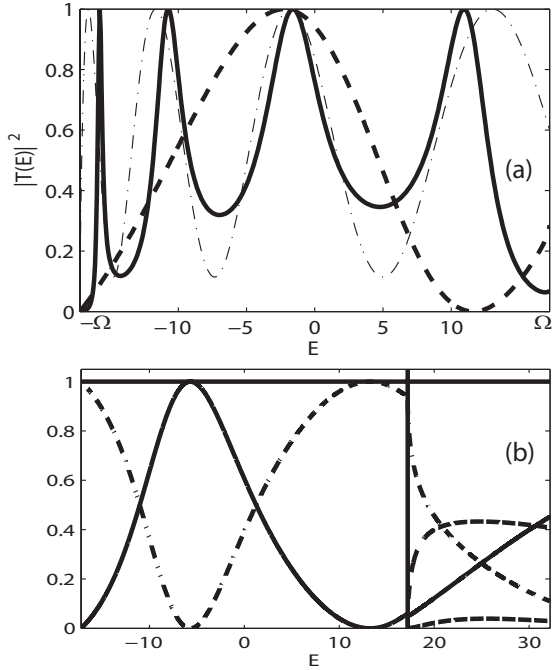}
  \caption{a) Transmission $|T(E)|^2$ in a $0$-$\pi$-$0$ junction versus energy  $-\Omega<E<\Omega$ for a left-incoming plane wave with $\Omega_0=14$, detuning $\delta=10$, $\pi$-zone length $a_1=0.4$ (dashed line), $a_2=2$ (solid line) and the square well approximation $\left|T_{sqw}(E;a_2)\right|^2$ (thin dashed line). 
  b) Transmission $|T(E)|^2$ (solid line) and reflection $|R(E)|^2$ (dashed dotted line) versus energy  for $a=0.5$. Above the energy border $E=\Omega$ (solid vertical line), we depict also excited channels transmission $|S_{1+,1-}|^2$ (dashed line) and $|S_{3+,1-}|^2$ (long dashed line). The constant line at $1$ (solid line) proves unitarity for all energies.}
  \label{fig:0pi0TRall}
\end{figure}

If we consider the scattering solutions for energies $-\Omega<E<\Omega$, then
the lower dressed state is an oscillatory and the excited component is 
an exponentially decaying state. 
Thus, we define the transmission amplitude $T(E)\equiv S_{3-,1-}$ 
as the forward scattering amplitude  for a left-incoming wave 
$\varphi^{in}=(0,1,0,0)$. This energy-dependent transmission is 
shown in Fig.~\ref{fig:0pi0TRall} a for two different lengths of the $\pi$-junction.
One observes the  typical transmission behavior with vanishing or low 
transmission at both sides of the energy range and resonances in between. 
It is intuitively clear, that there are more resonances with increasing junction length. 
This feature can be explained from an in-depth mathematical analysis of the poles of the S-matrix, or a qualitative physical reasoning.  

Indeed assuming an oscillatory solution in the lower dressed state manifold between the $\pi$-junction walls suggests the condition $\cos(k_-(E)a)=0$, like in a square-well. This analogy leads to several discrete resonances at the energies  $E_n=(n+1/2)^2\pi^2/a^2-\Omega$. With this approximation, we find the elementary but analytical expression 
\begin{align}\label{approx}
  \left|T_{sqw}(E)\right|^2=\cos^2(k_-a)\cos^4(2\xi)+\sin^2(k_- a)
\end{align}
for the transmission coefficient.
This square-well approximation  $\left|T_{sqw}(E;a_2)\right|^2$ is depicted for the length $a_2$ in Fig.~\ref{fig:0pi0TRall} a with a thin dashed line. It does match the exact solution quite well and reproduces the resonances up to some minor energy shifts.  

If we lift the limitations on scattering energies $E>\Omega$, then all four collision channels are energetically accessible and will be occupied. This situation is depicted in Fig.~\ref{fig:0pi0TRall} b, where we scan the whole energy range for a short $\pi$-junction length.  With the incoming state $\varphi^{in}\equiv (0,1,0,0)$, we
get for the outgoing amplitudes $\varphi^{out}=(S_{3+,1-},T=S_{3-,1-},S_{1+,1-},R=S_{1-,1-})$.
They satisfy the current conservation rule known as unitarity condition
\begin{equation}	|T|^2+|R|^2+\Re{\left(\frac{k_+}{k_-}\right)}(|S_{3+,1-}|^2+|S_{1+,1-}|^2)=1.
\end{equation}

In this Letter, we have provided an atomic model of  a $0$-$\pi$-$0$ Josephson junction, today realized with superconductors. We have studied a linear two-component Schr{\"o}dinger equation for a bosonic atomic gas, which is coupled by a phase-flipping laser field. On the mean-field level, this model demonstrates the emergence of macroscopic energy-degenerate quantum states, which are topologically distinct from flat-phase states. Their domain of existence is studied with phase-diagrams and as a function of the external system parameters, such as  the $\pi$- junction length $a$, the Rabi-frequency $\Omega_0$, or the detuning $\delta$. By further quantizing the Schr{\"o}dinger field, one can study the quantum evolution of these macroscopic energy-degenerate states, like quantum-  and thermally induced tunneling, or coherent oscillations, eventually.
 
{\bf Acknowledgment:} We gratefully acknowledge financial
support by the SFB/TRR 21 {\em ``Control of quantum correlations in tailored matter''} funded by the Deutsche Forschungsgemeinschaft (DFG). R.W. thanks the Deutsche Luft- und Raumfahrtagentur (DLR) for support from grant (50WM 1137).

\bibliographystyle{apsrev}
\bibliography{textbooks,bec,MyPublications,SF}

\hyphenation{Post-Script Sprin-ger}
\begin{thebibliography}{38}
\expandafter\ifx\csname natexlab\endcsname\relax\def\natexlab#1{#1}\fi
\expandafter\ifx\csname bibnamefont\endcsname\relax
  \def\bibnamefont#1{#1}\fi
\expandafter\ifx\csname bibfnamefont\endcsname\relax
  \def\bibfnamefont#1{#1}\fi
\expandafter\ifx\csname citenamefont\endcsname\relax
  \def\citenamefont#1{#1}\fi
\expandafter\ifx\csname url\endcsname\relax
  \def\url#1{\texttt{#1}}\fi
\expandafter\ifx\csname urlprefix\endcsname\relax\def\urlprefix{URL }\fi
\providecommand{\bibinfo}[2]{#2}
\providecommand{\eprint}[2][]{\url{#2}}

\bibitem[{\citenamefont{Debnath}(2004)}]{Debnath2004}
\bibinfo{author}{\bibfnamefont{L.}~\bibnamefont{Debnath}},
  \emph{\bibinfo{title}{Nonlinear Partial Differential Equations for Scientists
  and Engineers}} (\bibinfo{publisher}{Birkhäuser}, \bibinfo{address}{Boston},
  \bibinfo{year}{2004}).

\bibitem[{\citenamefont{Scott}(1969)}]{Scott1969}
\bibinfo{author}{\bibfnamefont{A.}~\bibnamefont{Scott}}, \bibinfo{journal}{Am.
  J. Phys.} \textbf{\bibinfo{volume}{37}}, \bibinfo{pages}{52}
  (\bibinfo{year}{1969}).

\bibitem[{\citenamefont{Barone et~al.}(1971)\citenamefont{Barone, Esposito,
  Magee, and Scott}}]{barone71}
\bibinfo{author}{\bibfnamefont{A.}~\bibnamefont{Barone}},
  \bibinfo{author}{\bibfnamefont{F.}~\bibnamefont{Esposito}},
  \bibinfo{author}{\bibfnamefont{C.}~\bibnamefont{Magee}}, \bibnamefont{and}
  \bibinfo{author}{\bibfnamefont{A.}~\bibnamefont{Scott}},
  \bibinfo{journal}{Nuovo Cimento} \textbf{\bibinfo{volume}{1}},
  \bibinfo{pages}{227} (\bibinfo{year}{1971}).

\bibitem[{\citenamefont{McCall and Hahn}(1969)}]{McCall1969}
\bibinfo{author}{\bibfnamefont{S.}~\bibnamefont{McCall}} \bibnamefont{and}
  \bibinfo{author}{\bibfnamefont{E.}~\bibnamefont{Hahn}},
  \bibinfo{journal}{Phys. Rev.} \textbf{\bibinfo{volume}{183}},
  \bibinfo{pages}{457} (\bibinfo{year}{1969}).

\bibitem[{\citenamefont{Lamb}(1971)}]{G_Lamb71}
\bibinfo{author}{\bibfnamefont{G.~L.} \bibnamefont{Lamb}},
  \bibinfo{journal}{Rev. Mod. Phys.} \textbf{\bibinfo{volume}{43}},
  \bibinfo{pages}{99} (\bibinfo{year}{1971}).

\bibitem[{\citenamefont{Mandelstam}(1975)}]{Mandelstam1975}
\bibinfo{author}{\bibfnamefont{S.}~\bibnamefont{Mandelstam}},
  \bibinfo{journal}{Phys. Rev. D} \textbf{\bibinfo{volume}{11}},
  \bibinfo{pages}{3026} (\bibinfo{year}{1975}).

\bibitem[{\citenamefont{Barone and Paterno}(1982)}]{barone82}
\bibinfo{author}{\bibfnamefont{A.}~\bibnamefont{Barone}} \bibnamefont{and}
  \bibinfo{author}{\bibfnamefont{G.}~\bibnamefont{Paterno}},
  \emph{\bibinfo{title}{Physics and Application of the Josephson Effect}}
  (\bibinfo{publisher}{Wiley Interscience}, \bibinfo{address}{New York},
  \bibinfo{year}{1982}).

\bibitem[{\citenamefont{Likharev}(1986)}]{likharev86}
\bibinfo{author}{\bibfnamefont{K.~K.} \bibnamefont{Likharev}},
  \emph{\bibinfo{title}{Dynamics of Josephson Junctions and Circuits}}
  (\bibinfo{publisher}{Gordon and Breach Publishers}, \bibinfo{year}{1986}).

\bibitem[{\citenamefont{Smerzi et~al.}(1997)\citenamefont{Smerzi, Fantoni,
  Giovanazzi, and Shenoy}}]{smerzi1097}
\bibinfo{author}{\bibfnamefont{A.}~\bibnamefont{Smerzi}},
  \bibinfo{author}{\bibfnamefont{S.}~\bibnamefont{Fantoni}},
  \bibinfo{author}{\bibfnamefont{S.}~\bibnamefont{Giovanazzi}},
  \bibnamefont{and} \bibinfo{author}{\bibfnamefont{S.~R.}
  \bibnamefont{Shenoy}}, \bibinfo{journal}{Phys.~Rev.~Lett.}
  \textbf{\bibinfo{volume}{79}}, \bibinfo{pages}{4950} (\bibinfo{year}{1997}).

\bibitem[{\citenamefont{Williams et~al.}(1999)\citenamefont{Williams, Walser,
  Cooper, Cornell, and Holland}}]{williams199}
\bibinfo{author}{\bibfnamefont{J.}~\bibnamefont{Williams}},
  \bibinfo{author}{\bibfnamefont{R.}~\bibnamefont{Walser}},
  \bibinfo{author}{\bibfnamefont{J.}~\bibnamefont{Cooper}},
  \bibinfo{author}{\bibfnamefont{E.}~\bibnamefont{Cornell}}, \bibnamefont{and}
  \bibinfo{author}{\bibfnamefont{M.}~\bibnamefont{Holland}},
  \bibinfo{journal}{Phys. Rev. A} \textbf{\bibinfo{volume}{59}},
  \bibinfo{pages}{R31} (\bibinfo{year}{1999}).

\bibitem[{\citenamefont{Leggett}(2001)}]{leggett401}
\bibinfo{author}{\bibfnamefont{A.}~\bibnamefont{Leggett}},
  \bibinfo{journal}{Rev. Mod. Phys.} \textbf{\bibinfo{volume}{73}},
  \bibinfo{pages}{307} (\bibinfo{year}{2001}).

\bibitem[{\citenamefont{Albiez et~al.}(2005)\citenamefont{Albiez, Gati,
  F{\"o}lling, Hunsmann, Cristian, and Oberthaler}}]{oberthaler05}
\bibinfo{author}{\bibfnamefont{M.}~\bibnamefont{Albiez}},
  \bibinfo{author}{\bibfnamefont{R.}~\bibnamefont{Gati}},
  \bibinfo{author}{\bibfnamefont{J.}~\bibnamefont{F{\"o}lling}},
  \bibinfo{author}{\bibfnamefont{S.}~\bibnamefont{Hunsmann}},
  \bibinfo{author}{\bibfnamefont{M.}~\bibnamefont{Cristian}}, \bibnamefont{and}
  \bibinfo{author}{\bibfnamefont{M.}~\bibnamefont{Oberthaler}},
  \bibinfo{journal}{Phys.~Rev.~Lett.} \textbf{\bibinfo{volume}{95}},
  \bibinfo{pages}{010402} (\bibinfo{year}{2005}).

\bibitem[{\citenamefont{Kaurov and Kuklov}(2005)}]{Kaurov2005}
\bibinfo{author}{\bibfnamefont{V.~M.} \bibnamefont{Kaurov}} \bibnamefont{and}
  \bibinfo{author}{\bibfnamefont{A.~B.} \bibnamefont{Kuklov}},
  \bibinfo{journal}{Phys.~Rev.~A} \textbf{\bibinfo{volume}{71}},
  \bibinfo{eid}{011601} (\bibinfo{year}{2005}).

\bibitem[{\citenamefont{Kaurov and Kuklov}(2006)}]{Kaurov2006}
\bibinfo{author}{\bibfnamefont{V.~M.} \bibnamefont{Kaurov}} \bibnamefont{and}
  \bibinfo{author}{\bibfnamefont{A.~B.} \bibnamefont{Kuklov}},
  \bibinfo{journal}{Phys.~Rev.~A} \textbf{\bibinfo{volume}{73}},
  \bibinfo{eid}{013627} (\bibinfo{year}{2006}).

\bibitem[{\citenamefont{Brand et~al.}(2009)\citenamefont{Brand, Haigh, and
  Z\"ulicke}}]{brand09}
\bibinfo{author}{\bibfnamefont{J.}~\bibnamefont{Brand}},
  \bibinfo{author}{\bibfnamefont{T.~J.} \bibnamefont{Haigh}}, \bibnamefont{and}
  \bibinfo{author}{\bibfnamefont{U.}~\bibnamefont{Z\"ulicke}},
  \bibinfo{journal}{Phys. Rev. A} \textbf{\bibinfo{volume}{80}},
  \bibinfo{pages}{011602} (\bibinfo{year}{2009}).

\bibitem[{\citenamefont{Ramanathan et~al.}(2011)\citenamefont{Ramanathan,
  Wright, Muniz, Zelan, Hill, Lobb, Helmerson, Phillips, and
  Campbell}}]{phillips11}
\bibinfo{author}{\bibfnamefont{A.}~\bibnamefont{Ramanathan}},
  \bibinfo{author}{\bibfnamefont{K.~C.} \bibnamefont{Wright}},
  \bibinfo{author}{\bibfnamefont{S.~R.} \bibnamefont{Muniz}},
  \bibinfo{author}{\bibfnamefont{M.}~\bibnamefont{Zelan}},
  \bibinfo{author}{\bibfnamefont{W.~T.} \bibnamefont{Hill}},
  \bibinfo{author}{\bibfnamefont{C.~J.} \bibnamefont{Lobb}},
  \bibinfo{author}{\bibfnamefont{K.}~\bibnamefont{Helmerson}},
  \bibinfo{author}{\bibfnamefont{W.~D.} \bibnamefont{Phillips}},
  \bibnamefont{and} \bibinfo{author}{\bibfnamefont{G.~K.}
  \bibnamefont{Campbell}}, \bibinfo{journal}{Phys. Rev. Lett.}
  \textbf{\bibinfo{volume}{106}}, \bibinfo{pages}{130401}
  (\bibinfo{year}{2011}).

\bibitem[{\citenamefont{Kivshar and Malomed}(1989)}]{kivshar89}
\bibinfo{author}{\bibfnamefont{Y.~S.} \bibnamefont{Kivshar}} \bibnamefont{and}
  \bibinfo{author}{\bibfnamefont{B.~A.} \bibnamefont{Malomed}},
  \bibinfo{journal}{Rev. Mod. Phys.} \textbf{\bibinfo{volume}{61}},
  \bibinfo{pages}{763} (\bibinfo{year}{1989}).

\bibitem[{\citenamefont{Bulaevskii et~al.}(1978)\citenamefont{Bulaevskii,
  Kuzii, and Sobyanin}}]{Bulaevskii:0-pi-LJJ}
\bibinfo{author}{\bibfnamefont{L.~N.} \bibnamefont{Bulaevskii}},
  \bibinfo{author}{\bibfnamefont{V.~V.} \bibnamefont{Kuzii}}, \bibnamefont{and}
  \bibinfo{author}{\bibfnamefont{A.~A.} \bibnamefont{Sobyanin}},
  \bibinfo{journal}{Solid State Commun.} \textbf{\bibinfo{volume}{25}},
  \bibinfo{pages}{1053} (\bibinfo{year}{1978}).

\bibitem[{\citenamefont{Xu et~al.}(1995)\citenamefont{Xu, Miller, and
  Ting}}]{Xu:SF-shape}
\bibinfo{author}{\bibfnamefont{J.~H.} \bibnamefont{Xu}},
  \bibinfo{author}{\bibfnamefont{J.~H.} \bibnamefont{Miller}},
  \bibnamefont{and} \bibinfo{author}{\bibfnamefont{C.~S.} \bibnamefont{Ting}},
  \bibinfo{journal}{Phys. Rev. B} \textbf{\bibinfo{volume}{51}},
  \bibinfo{pages}{11958} (\bibinfo{year}{1995}).

\bibitem[{\citenamefont{Kirtley et~al.}(1996)\citenamefont{Kirtley, Tsuei,
  Rupp, Sun, Yu-Jahnes, Gupta, Ketchen, Moler, and Bhushan}}]{Kirtley:SF:HTSGB}
\bibinfo{author}{\bibfnamefont{J.~R.} \bibnamefont{Kirtley}},
  \bibinfo{author}{\bibfnamefont{C.~C.} \bibnamefont{Tsuei}},
  \bibinfo{author}{\bibfnamefont{M.}~\bibnamefont{Rupp}},
  \bibinfo{author}{\bibfnamefont{J.~Z.} \bibnamefont{Sun}},
  \bibinfo{author}{\bibfnamefont{L.~S.} \bibnamefont{Yu-Jahnes}},
  \bibinfo{author}{\bibfnamefont{A.}~\bibnamefont{Gupta}},
  \bibinfo{author}{\bibfnamefont{M.~B.} \bibnamefont{Ketchen}},
  \bibinfo{author}{\bibfnamefont{K.~A.} \bibnamefont{Moler}}, \bibnamefont{and}
  \bibinfo{author}{\bibfnamefont{M.}~\bibnamefont{Bhushan}},
  \bibinfo{journal}{Phys. Rev. Lett.} \textbf{\bibinfo{volume}{76}},
  \bibinfo{pages}{1336} (\bibinfo{year}{1996}).

\bibitem[{\citenamefont{Hilgenkamp et~al.}(2003)\citenamefont{Hilgenkamp,
  Ariando, Smilde, Blank, Rijnders, Rogalla, Kirtley, and
  Tsuei}}]{Hilgenkamp:zigzag:SF}
\bibinfo{author}{\bibfnamefont{H.}~\bibnamefont{Hilgenkamp}},
  \bibinfo{author}{\bibnamefont{Ariando}},
  \bibinfo{author}{\bibfnamefont{H.-J.~H.} \bibnamefont{Smilde}},
  \bibinfo{author}{\bibfnamefont{D.~H.~A.} \bibnamefont{Blank}},
  \bibinfo{author}{\bibfnamefont{G.}~\bibnamefont{Rijnders}},
  \bibinfo{author}{\bibfnamefont{H.}~\bibnamefont{Rogalla}},
  \bibinfo{author}{\bibfnamefont{J.~R.} \bibnamefont{Kirtley}},
  \bibnamefont{and} \bibinfo{author}{\bibfnamefont{C.~C.} \bibnamefont{Tsuei}},
  \bibinfo{journal}{Nature} \textbf{\bibinfo{volume}{422}}, \bibinfo{pages}{50}
  (\bibinfo{year}{2003}).

\bibitem[{\citenamefont{Goldobin et~al.}(2002)\citenamefont{Goldobin, Koelle,
  and Kleiner}}]{Goldobin:SF-Shape}
\bibinfo{author}{\bibfnamefont{E.}~\bibnamefont{Goldobin}},
  \bibinfo{author}{\bibfnamefont{D.}~\bibnamefont{Koelle}}, \bibnamefont{and}
  \bibinfo{author}{\bibfnamefont{R.}~\bibnamefont{Kleiner}},
  \bibinfo{journal}{Phys. Rev. B} \textbf{\bibinfo{volume}{66}},
  \bibinfo{pages}{100508(R)} (\bibinfo{year}{2002}).

\bibitem[{\citenamefont{Nappi et~al.}(2004)\citenamefont{Nappi, Lisitskiy,
  Rotoli, Cristiano, and Barone}}]{nappi04}
\bibinfo{author}{\bibfnamefont{C.}~\bibnamefont{Nappi}},
  \bibinfo{author}{\bibfnamefont{M.~P.} \bibnamefont{Lisitskiy}},
  \bibinfo{author}{\bibfnamefont{G.}~\bibnamefont{Rotoli}},
  \bibinfo{author}{\bibfnamefont{R.}~\bibnamefont{Cristiano}},
  \bibnamefont{and} \bibinfo{author}{\bibfnamefont{A.}~\bibnamefont{Barone}},
  \bibinfo{journal}{Phys. Rev. Lett.} \textbf{\bibinfo{volume}{93}},
  \bibinfo{pages}{187001} (\bibinfo{year}{2004}).

\bibitem[{\citenamefont{Goldobin et~al.}(2005)\citenamefont{Goldobin, Vogel,
  Crasser, Walser, Schleich, Koelle, and Kleiner}}]{goldobin05}
\bibinfo{author}{\bibfnamefont{E.}~\bibnamefont{Goldobin}},
  \bibinfo{author}{\bibfnamefont{K.}~\bibnamefont{Vogel}},
  \bibinfo{author}{\bibfnamefont{O.}~\bibnamefont{Crasser}},
  \bibinfo{author}{\bibfnamefont{R.}~\bibnamefont{Walser}},
  \bibinfo{author}{\bibfnamefont{W.~P.} \bibnamefont{Schleich}},
  \bibinfo{author}{\bibfnamefont{D.}~\bibnamefont{Koelle}}, \bibnamefont{and}
  \bibinfo{author}{\bibfnamefont{R.}~\bibnamefont{Kleiner}},
  \bibinfo{journal}{Phys. Rev. B} \textbf{\bibinfo{volume}{72}},
  \bibinfo{pages}{054527} (\bibinfo{year}{2005}).

\bibitem[{\citenamefont{Vogel et~al.}(2009)\citenamefont{Vogel, Schleich, Kato,
  Koelle, Kleiner, and Goldobin}}]{vogel09}
\bibinfo{author}{\bibfnamefont{K.}~\bibnamefont{Vogel}},
  \bibinfo{author}{\bibfnamefont{W.~P.} \bibnamefont{Schleich}},
  \bibinfo{author}{\bibfnamefont{T.}~\bibnamefont{Kato}},
  \bibinfo{author}{\bibfnamefont{D.}~\bibnamefont{Koelle}},
  \bibinfo{author}{\bibfnamefont{R.}~\bibnamefont{Kleiner}}, \bibnamefont{and}
  \bibinfo{author}{\bibfnamefont{E.}~\bibnamefont{Goldobin}},
  \bibinfo{journal}{Phys. Rev. B} \textbf{\bibinfo{volume}{80}},
  \bibinfo{pages}{134515} (\bibinfo{year}{2009}).

\bibitem[{\citenamefont{Goldobin et~al.}(2010)\citenamefont{Goldobin, Vogel,
  Schleich, Koelle, and Kleiner}}]{goldobin10}
\bibinfo{author}{\bibfnamefont{E.}~\bibnamefont{Goldobin}},
  \bibinfo{author}{\bibfnamefont{K.}~\bibnamefont{Vogel}},
  \bibinfo{author}{\bibfnamefont{W.~P.} \bibnamefont{Schleich}},
  \bibinfo{author}{\bibfnamefont{D.}~\bibnamefont{Koelle}}, \bibnamefont{and}
  \bibinfo{author}{\bibfnamefont{R.}~\bibnamefont{Kleiner}},
  \bibinfo{journal}{Phys. Rev. B} \textbf{\bibinfo{volume}{81}},
  \bibinfo{pages}{054514} (\bibinfo{year}{2010}).

\bibitem[{\citenamefont{Hall et~al.}(1998)\citenamefont{Hall, Matthews, Ensher,
  Wieman, and Cornell}}]{hall98}
\bibinfo{author}{\bibfnamefont{D.~S.} \bibnamefont{Hall}},
  \bibinfo{author}{\bibfnamefont{M.~R.} \bibnamefont{Matthews}},
  \bibinfo{author}{\bibfnamefont{J.~R.} \bibnamefont{Ensher}},
  \bibinfo{author}{\bibfnamefont{C.~E.} \bibnamefont{Wieman}},
  \bibnamefont{and} \bibinfo{author}{\bibfnamefont{E.~A.}
  \bibnamefont{Cornell}}, \bibinfo{journal}{Phys. Rev. Lett.}
  \textbf{\bibinfo{volume}{81}}, \bibinfo{pages}{1539} (\bibinfo{year}{1998}).

\bibitem[{\citenamefont{Kuklov and Birman}(2000)}]{kuklov00}
\bibinfo{author}{\bibfnamefont{A.~B.} \bibnamefont{Kuklov}} \bibnamefont{and}
  \bibinfo{author}{\bibfnamefont{J.~L.} \bibnamefont{Birman}},
  \bibinfo{journal}{Phys. Rev. Lett.} \textbf{\bibinfo{volume}{85}},
  \bibinfo{pages}{5488} (\bibinfo{year}{2000}).

\bibitem[{\citenamefont{Harber et~al.}(2002)\citenamefont{Harber, Lewandowski,
  McGuirk, and Cornell}}]{lewandowski02}
\bibinfo{author}{\bibfnamefont{D.~M.} \bibnamefont{Harber}},
  \bibinfo{author}{\bibfnamefont{H.~J.} \bibnamefont{Lewandowski}},
  \bibinfo{author}{\bibfnamefont{J.~M.} \bibnamefont{McGuirk}},
  \bibnamefont{and} \bibinfo{author}{\bibfnamefont{E.~A.}
  \bibnamefont{Cornell}}, \bibinfo{journal}{Phys. Rev. A}
  \textbf{\bibinfo{volume}{66}}, \bibinfo{pages}{053616}
  (\bibinfo{year}{2002}).

\bibitem[{\citenamefont{Walser et~al.}(2008)\citenamefont{Walser, Goldobin,
  Crasser, Koelle, Kleiner, and Schleich}}]{walser07}
\bibinfo{author}{\bibfnamefont{R.}~\bibnamefont{Walser}},
  \bibinfo{author}{\bibfnamefont{E.}~\bibnamefont{Goldobin}},
  \bibinfo{author}{\bibfnamefont{O.}~\bibnamefont{Crasser}},
  \bibinfo{author}{\bibfnamefont{D.}~\bibnamefont{Koelle}},
  \bibinfo{author}{\bibfnamefont{R.}~\bibnamefont{Kleiner}}, \bibnamefont{and}
  \bibinfo{author}{\bibfnamefont{W.~P.} \bibnamefont{Schleich}},
  \bibinfo{journal}{New J. Phys.} \textbf{\bibinfo{volume}{10}},
  \bibinfo{pages}{45020} (\bibinfo{year}{2008}).

\bibitem[{foo()}]{footnote120911}
\bibinfo{note}{Dimensionless units are assumed with $\hbar=2m=1$.}

\bibitem[{\citenamefont{Schleich}(2001)}]{schleich01}
\bibinfo{author}{\bibfnamefont{W.~P.} \bibnamefont{Schleich}},
  \emph{\bibinfo{title}{Quantum Optics in Phase Space}}
  (\bibinfo{publisher}{Wiley-VCH}, \bibinfo{address}{Weinheim},
  \bibinfo{year}{2001}).

\bibitem[{\citenamefont{Dum et~al.}(1998)\citenamefont{Dum, Cirac, Lewenstein,
  and Zoller}}]{Dum1998a}
\bibinfo{author}{\bibfnamefont{R.}~\bibnamefont{Dum}},
  \bibinfo{author}{\bibfnamefont{J.~I.} \bibnamefont{Cirac}},
  \bibinfo{author}{\bibfnamefont{M.}~\bibnamefont{Lewenstein}},
  \bibnamefont{and} \bibinfo{author}{\bibfnamefont{P.}~\bibnamefont{Zoller}},
  \bibinfo{journal}{Phys.~Rev.~Lett.} \textbf{\bibinfo{volume}{80}},
  \bibinfo{pages}{2972} (\bibinfo{year}{1998}).

\bibitem[{\citenamefont{Dobrek et~al.}(1999)\citenamefont{Dobrek, Gajda,
  Lewenstein, Sengstock, Birkl, and Ertmer}}]{dobrek99}
\bibinfo{author}{\bibfnamefont{L.}~\bibnamefont{Dobrek}},
  \bibinfo{author}{\bibfnamefont{M.}~\bibnamefont{Gajda}},
  \bibinfo{author}{\bibfnamefont{M.}~\bibnamefont{Lewenstein}},
  \bibinfo{author}{\bibfnamefont{K.}~\bibnamefont{Sengstock}},
  \bibinfo{author}{\bibfnamefont{G.}~\bibnamefont{Birkl}}, \bibnamefont{and}
  \bibinfo{author}{\bibfnamefont{W.}~\bibnamefont{Ertmer}},
  \bibinfo{journal}{Phys.~Rev.~A} \textbf{\bibinfo{volume}{60}},
  \bibinfo{pages}{R3381} (\bibinfo{year}{1999}).

\bibitem[{\citenamefont{Denschlag et~al.}(2000)\citenamefont{Denschlag,
  Simsarian, Feder, Clark, Collins, Cubizolles, Deng, Hagley, Helmerson,
  Reinhardt et~al.}}]{denschlag00}
\bibinfo{author}{\bibfnamefont{J.}~\bibnamefont{Denschlag}},
  \bibinfo{author}{\bibfnamefont{J.~E.} \bibnamefont{Simsarian}},
  \bibinfo{author}{\bibfnamefont{D.~L.} \bibnamefont{Feder}},
  \bibinfo{author}{\bibfnamefont{C.~W.} \bibnamefont{Clark}},
  \bibinfo{author}{\bibfnamefont{L.~A.} \bibnamefont{Collins}},
  \bibinfo{author}{\bibfnamefont{J.}~\bibnamefont{Cubizolles}},
  \bibinfo{author}{\bibfnamefont{L.}~\bibnamefont{Deng}},
  \bibinfo{author}{\bibfnamefont{E.~W.} \bibnamefont{Hagley}},
  \bibinfo{author}{\bibfnamefont{K.}~\bibnamefont{Helmerson}},
  \bibinfo{author}{\bibfnamefont{W.~P.} \bibnamefont{Reinhardt}},
  \bibnamefont{et~al.}, \bibinfo{journal}{Science}
  \textbf{\bibinfo{volume}{287}}, \bibinfo{pages}{97} (\bibinfo{year}{2000}).

\bibitem[{\citenamefont{Strecker et~al.}(2002)\citenamefont{Strecker,
  Partridge, Truscott, and Hulet}}]{strecker502}
\bibinfo{author}{\bibfnamefont{K.}~\bibnamefont{Strecker}},
  \bibinfo{author}{\bibfnamefont{G.}~\bibnamefont{Partridge}},
  \bibinfo{author}{\bibfnamefont{A.}~\bibnamefont{Truscott}}, \bibnamefont{and}
  \bibinfo{author}{\bibfnamefont{R.}~\bibnamefont{Hulet}},
  \bibinfo{journal}{Nature} \textbf{\bibinfo{volume}{417}},
  \bibinfo{pages}{150} (\bibinfo{year}{2002}).

\bibitem[{\citenamefont{Kazantsev et~al.}(1990)\citenamefont{Kazantsev,
  Surdutovich, and Yakovlev}}]{kazantsev90}
\bibinfo{author}{\bibfnamefont{A.}~\bibnamefont{Kazantsev}},
  \bibinfo{author}{\bibfnamefont{G.}~\bibnamefont{Surdutovich}},
  \bibnamefont{and} \bibinfo{author}{\bibfnamefont{V.}~\bibnamefont{Yakovlev}},
  \emph{\bibinfo{title}{Mechanical Action of Light on Atoms}}
  (\bibinfo{publisher}{World Scientific}, \bibinfo{address}{Singapore},
  \bibinfo{year}{1990}).

\bibitem[{\citenamefont{Fetter}(2002)}]{fetter2002}
\bibinfo{author}{\bibfnamefont{A.}~\bibnamefont{Fetter}},
  \bibinfo{journal}{JLTP} \textbf{\bibinfo{volume}{129}}, \bibinfo{pages}{263}
  (\bibinfo{year}{2002}).

\end{thebibliography}
\end{document}